\documentstyle[aps,multicol,epsf]{revtex}

\begin{document}
\draft

\title{Kinematic Segregation of Flowing Grains in Sandpiles}
\author{Hern\'an A. Makse \footnote{Present address: Schlumberger-Doll 
Research, Old Quarry Road, Ridgefield, CT 06877, USA}}
 
\address{
Laboratoire de Physique de la Mati\`ere Condens\'ee, Coll\`ege de
France\\ 
11 place Marcelin Berthelot, 75231 Paris Cedex 05, France, \\
and  Center for Polymer Studies, and Physics Dept., Boston
University, Boston, MA 02215 USA}
\date{submitted to Eur. Phys. J.-B}
\maketitle
\begin{abstract}
We study the segregation of granular mixtures in  two-dimensional silos
using a set of coupled equations for surface flows 
of grains. We study the thick flow regime, where the grains are
segregated
in the rolling phase. We incorporate this dynamical segregation process,
called kinematic sieving, free-surface segregation or percolation,
into the 
theoretical formalism and calculate the profiles of the rolling species 
and the concentration of grains  in the bulk in the steady state.
Our solution shows the segregation of the mixture with the large grains
being found at the bottom of the pile in qualitative 
agreement with experiments.

\end{abstract}

\begin{multicols}{2}

\narrowtext

\section{Introduction}

\begin{figure}
\centerline{
\vbox{ \hbox{\epsfxsize=8.cm
\epsfbox{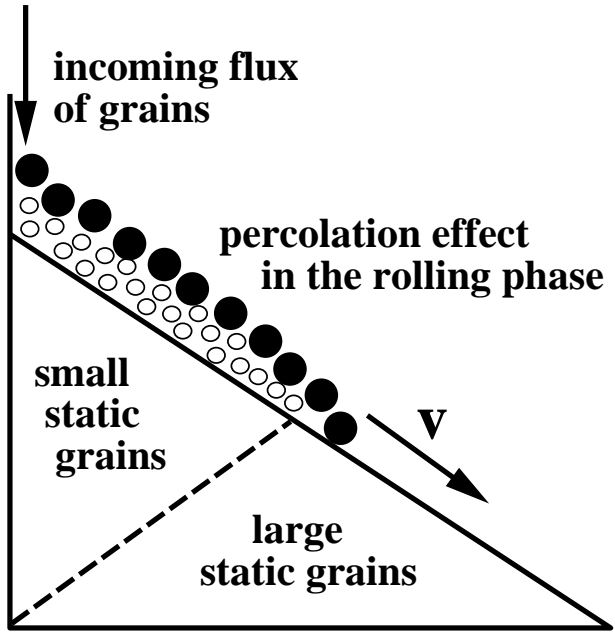}}
}}
\vspace{1cm}
\narrowtext
\caption{ Schematic representation of the percolation effect 
in the rolling phase. We pour continuously a mixture of large and small 
grain in a two-dimensional cell and a steady thick flow of grains is observed.
The percolation effect consists on the size segregation of grains in the
 rolling phase: large grains are observed to rise to the top of the 
rolling phase,
and small grains drifts to the bottom of the rolling phase. Due to the 
percolation effect, the small grains are the first to be captured at the 
pile surface resulting in the segregation of the mixture in the bulk.}
\label{perco}
\end{figure}

When a mixture of grains 
\cite{bagnold,review1,review2,segregation1,segregation2,segregation3}
differing in size and shape is poured in a
two-dimensional cell consisting of two parallel vertical slabs separated
by a narrow  gap (i.e., a granular Hele-Shaw cell) the grains segregate
according to their size and shape.
The typical experimental set
consists of a vertical
``quasi-two-dimensional'' Hele-Shaw cell  with a narrow gap
of 5 mm separating two transparent plates of 300 mm by 200 mm (see Fig.
\ref{perco}).  We close the left edge of the cell leaving
the right edge free, and we pour continuously, near the left edge,
 an equal-volume
mixture of grains differing in size and shape.
When the larger grains are more rounded than the small grains, a
segregation 
of the mixture is observed with the large grains being found at the
bottom of the cell and the small at the top.
When the large grains are rougher than the small grains a periodic 
pattern consisting of layers of large and small grains parallel to the
pile surface occurs \cite{makse1,makse3,makse2,cms,makse4,yan,kaka}. 
This periodic pattern is called spontaneous 
stratification and it was found to
occur for a wide range of size ratios between the grains:
$d_2/d_1 > 1.4$, where $d_1$ is the typical size of the small grains and
 $d_2$ is the typical size of the large grains.
When the size ratio between the grains is close to one, only segregation
is found no matter the shape of the grains, i.e. no matter the angle of
repose of the species. 

An important segregation 
mechanism acting when $d_2/d_1$ is not close to one, and
in the the case of thick flows is the so called kinematic sieving 
\cite{savage1}, free surface segregation \cite{drahun} or percolation effect
\cite{makse3,cms}. 
Due to this  phenomenon the 
large grains in the rolling phase
are found to rise to the top of the rolling phase while the small grains
 sink downward through the gaps left by
the motion of larger grains in the rolling phase (Fig. \ref{perco}).
Since only the small grains interact with the surface they are captured
at the surface first causing the
larger grains to be convected further to the bottom of the pile.
Thus, a strong segregation effect results in a pile
with the small grains being found at the top of the pile and the 
large grains being found 
at the bottom of the pile.
This effect has been observed in high-speed photography experiments
to be important for segregation and stratification of granular mixtures 
\cite{makse3} in well developed flows of granular mixtures.
 
In this paper, we study analytically the segregation 
of granular mixtures poured in two-dimensional sandpiles
due  percolation in the rolling phase 
by using a suitable modification to the 
equations of motion for the rolling species 
interacting with the  static grains of the pile.
We calculate the corrections to the steady-state profiles due to the
percolation effect and compare with theoretical predictions without
percolation effects and with experimental observations.

Steady state solutions for the granular flows of grains in two-dimensional
geometries have been calculated in \cite{pgg} for the case of a single
species flows in thin rotating drums, in \cite{bdg} for the case of
two-species differing in angle of repose and flowing in two-dimensional
 silos, and in \cite{makse2,makse4}
for  thin flows of grains in two-dimensional silos
with two species of grains differing 
 in size and shape.
When the grains differ  only in angle of repose, 
the solution shows the segregation of the mixture and the
concentrations decay slowly as a power-law of the position in the cell 
\cite{bdg}.
When there is a large difference in size and shape the solution shows
a strong segregation of the mixture and the profiles decay exponentially
fast \cite{makse2,makse4}. 
The segregation solution is only valid for mixtures of large-rounded grains
and small rough grains. Otherwise, the steady 
state solution is unstable, leading to the stratification of the mixture.

Here, we treat  the 
 case when the difference in  size of the species  is not too small 
(i.e. $d_2/d_1 > 1.4$) but, as opposed to  \cite{makse4},  
we focus on 
 the thick flow regime. In this regime the thickness of the rolling phase
is expected to be large, and it is possible to observe the 
percolation effect (the segregation in the rolling phase) \cite{makse3}.
The steady-state 
solution we find shows the complete segregation of the mixture with
the large grains being found at the bottom of the pile, and we compare 
this  solution with the solution corresponding to the case of thin flows
when the percolation effect is absent.

The paper is organized as follows: In Section \ref{theory} we present
the theoretical formalism 
for surface flows of granular mixtures. 
In Section  \ref{ssregime} we calculate the steady-state solution of the
problem,   discuss our results, and compare
to the theoretical predictions obtained without the percolation effect.

\section{Surface Flow of Granular Mixtures}
\label{theory}

The theoretical study of surface flows of granular materials was
triggered by the works of Bouchaud, Cates, R. Prakash, and Edwards
(BCRE)  \cite{bouchaud} and Mehta and
collaborators \cite{mehta}.
In a recent theoretical study for 
the case of a single-species sandpile
BCRE \cite{bouchaud}
proposed 
two coupled variables to describe
the  dynamics of two-dimensional sandpile surfaces:
the local angle of the sandpile $\theta(x,t)$ (or alternatively the
height of the sandpile $h(x,t)$) which describes the static phase
(i.e., the grains which belong to the pile),
 and the local thickness of the layer of
rolling grains $R(x,t)$ to describe the rolling phase (i.e., the grains
that are not part of the pile but roll downwards on top of the static
phase). 
BCRE also proposed a set of 
convective-diffusion
equation 
for the rolling grains,
which was later simplified by de Gennes \cite{pgg}.

Recently, Boutreux and de Gennes (BdG)
 \cite{bdg} have extended the BCRE formalism
to the case of two species.
This formalism considers the two local 
``equivalent thicknesses'' of the species in the rolling phase 
$R_{\alpha}(x, t)$ (i.e. the total  thickness of the rolling phase
 multiplied by the
local volume fraction of the $\alpha$ grains in the 
rolling phase at position $x$), with $\alpha =1,2$ respectively for small
and large grains.
The total thickness of the rolling phase is defined as 
\begin{equation}
R(x,t) \equiv
R_1(x,t) + R_2(x,t).
\end{equation}
The static phase is described by
 the height of the sandpile $h(x,t)$, and the volume
fraction of static grains $\phi_\alpha(x,t)$ of type $\alpha$ at the surface
of the pile.
We consider a silo
or cell of lateral size $L$
and the pouring point is assumed to be at $x=0$.
For notational convenience we do not consider the difference between the
angle and the tangent of the angle, i.e. 
$\theta(x,t) \equiv -\frac{\partial h}{\partial x}$.
The equations of motion for the rolling species are \cite{bdg}
 
\begin{mathletters}
\label{bdg-eq}
\begin{equation}
\label{bdg-r}
\frac{\partial R_\alpha(x,t)}{\partial t}=-v_\alpha \frac{\partial
R_\alpha}{\partial x} +
\Gamma_\alpha,
\end{equation}
and the equation for $h(x,t)$ follows by conservation
 
\begin{equation}
\label{bdg-h}
 \frac{\partial h(x,t)}{\partial t}=- 
\Gamma_1-\Gamma_2.
\end{equation}
\end{mathletters}

Here $v_\alpha$ is the downhill convection velocity of species
$\alpha$
along $x$, 
 assumed to be
constant in space and in time.
The interaction term $\Gamma_\alpha$  
takes into account the conversion of static grains into
rolling grains
and vice versa, and 
it is  defined as
\begin{mathletters}
\label{gamma}
\begin{equation}
\Gamma_1 
\equiv a_1(\theta) \phi_1 R_1 - b_1(\theta) R_1
\end{equation}
\begin{equation}
\Gamma_2 
\equiv a_2(\theta) \phi_2 R_2 e^{-\lambda R_1/R} - b_2(\theta) R_2  e^{-\lambda R_1/R} 
\end{equation}
\end{mathletters}

This definition involves 
a set of a priori unknown collision 
functions contributing to the rate processes:
$a_\alpha(\theta)$ is the contribution due to an amplification process,
(i.e., when a static grain of type $\alpha$ is converted into a rolling grain 
due to  
a collision by a rolling grain of type $\alpha$), $b_\alpha(\theta)$
is the contribution due to capture of a rolling grain of type $\alpha$,
(i.e., when a rolling grain of type $\alpha$ is converted into a static
grain).
Cross-amplification
processes, (i.e., the amplification of a static grain of type $\beta$ due to
a collision by a rolling grain of type $\alpha$) do not contribute to
the exchange of grains since these kind of processes are inhibited by the
percolation effect. Therefore, we do not consider them in the
definition of the interaction term.

The interaction between rolling and static grains is assumed to be proportional
to the number of interacting rolling grain, then it is proportional to 
$R_\alpha(x,t)$. This is clearly valid for thin flows. However, in the case of 
thick flows treated here, this approximation might
still be valid \cite{cms} since the interaction might be
proportional to the pressure exerted by the fluid phase 
which in turn is proportional to thickness of the rolling phase \cite{zik}.

Due to percolation effect the interaction with the bulk of the large
rolling grains $R_2(x,t)$ is greatly reduced as long as there are 
small rolling grain $R_1(x,t)$ interacting with the surface. Therefore, 
to take into account the percolation effect, 
we replace $R_2(x,t)$ in the definition of the interaction term
by  $R_2(x,t) \exp[\lambda R_1(x,t)/R(x,t)]$.
The exponential factor multiplying  $R_2(x,t)$ mimics the fact that 
the interaction of large grains is screened by the presence of small 
grains so that $R_2$ interact with the grain at the surface of the static 
phase only when $R_1(x,t)\ll R(x,t)/\lambda$. 
The dimensionless parameter
$\lambda>0$ measures the degree of percolation, with $\lambda = 0$ 
corresponding to the no 
percolation case treated in \cite{makse2}.

The concentrations  of static grains at the surface of the pile 
$\phi_\alpha(x,t)$ are given by
\begin{mathletters} 
\begin{equation}
\phi_\alpha(x,t) \frac{\partial h}{\partial t} = - \Gamma_\alpha,
\label{bdg-phi}
\end{equation}
and
\begin{equation}
 \phi_1 + \phi_2 =1.
\end{equation}
\end{mathletters} 

We use the following definitions of the collision functions
\cite{makse2}:

\begin{mathletters}
\label{canonical2}
\begin{equation}
\begin{array}{llcl}
a_\alpha(\theta)&\equiv& \gamma_{\alpha\alpha} & \Pi[\theta(x,t)-
\theta_\alpha(\phi_\beta)] \\
b_\alpha(\theta)&\equiv& \gamma_{\alpha\alpha} &
\Pi[\theta_\alpha(\phi_\beta)-\theta(x,t)] 
\end{array}
\end{equation}
where 
 
\begin{equation}
\Pi[x] \equiv \left \{
\begin{array}{cr}
0  &~~~~~~~\mbox{if $x < 0$} \\
x  &~~~~~~~\mbox{if $x \ge 0$}
\end{array}
\right . .
\end{equation}
\end{mathletters}

Here, the rates $\gamma_{\alpha\alpha} > 0$ has dimension of inverse time,
and $v_\alpha / \gamma_{\alpha\alpha} \sim d_\alpha$. 
The generalized  angle of repose
$\theta_\alpha(\phi_\beta)$ of a $\alpha$ type of rolling grain is a continuous
function of the composition of the surface
$\phi_\beta$ \cite{makse2}, and we 
also define 
$\theta_{\alpha\beta}$ as
$\theta_\alpha(\phi_\beta)$ for $\phi_\beta=1$

\begin{equation}
\label{dependence}
\begin{array}{rcl}
\theta_1(\phi_2) &=& m \phi_2 + \theta_{11}\\
\theta_2(\phi_2) &=& m \phi_2 + \theta_{21} = - m \phi_1 + \theta_{22},
\end{array}
\end{equation}
where $m\equiv \theta_{12} - \theta_{11} = \theta_{22} - \theta_{21}$.
We assume the difference 
$\psi \equiv
\theta_1(\phi_2)-\theta_2(\phi_2)$
 to be
independent of the concentration $\phi_2$.
We notice that 
for mixtures of grains with different shapes we have
$\theta_{11}\neq\theta_{22}$ 
(here $\theta_{11}$ is the angle of 
the pure small species and  $\theta_{22}$ denote 
the angle of repose of the pure large species), and different sizes
$\theta_{12}\neq\theta_{21}$. 
In the following we treat only the case when the type $1$
 small grains are the roughest
(i.e. $\theta_{22}<\theta_{11}$, since the angle of repose depends
on the surface properties of the grains and it is larger for the roughest
grains) or when the grains differ only in size (i.e.,  
$\theta_{22}=\theta_{11}$). The case when the large grains are the roughest 
 ($\theta_{11}<\theta_{22}$) leads to stratification of the mixture and it is 
treated
in \cite{makse2,cms,makse4}.

\begin{figure}
\centerline{
\vbox{ \hbox{\epsfxsize=8.cm
\epsfbox{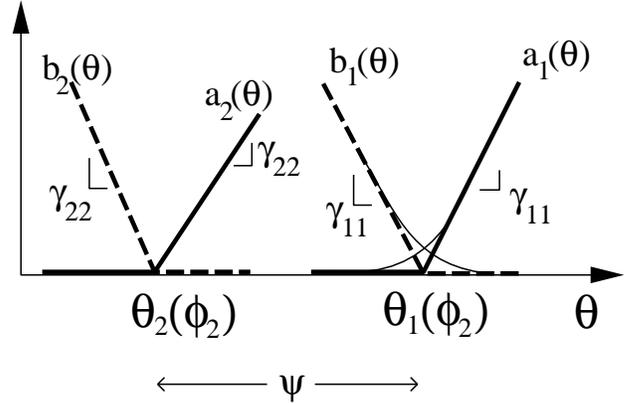}}
}}
\vspace{1cm}
\narrowtext
\caption{Plot of capture $b_\alpha(\theta)$ and
amplification  $a_\alpha(\theta)$
functions.
These functions are expected to be continuous in a region near
the angle of repose (as shown by the thin curves). However, when 
the 
the ratio between the size of the grains is $d_2/d_1 > 1.4$ then 
$\psi = \theta_1(\phi_2)- \theta_2(\phi_2)$ is large enough so 
we 
can approximate these
functions by the forms shown in this figure.}
\label{functions}
\end{figure}

The collision functions $a_\alpha$ and $b_\alpha$
are expected to be continuous in the region of interest near
the angle of repose (as shown, for instance, by the thin curves in
Fig. \ref{functions}). 
However, when the 
 ratio between the size of the grains is large enough, 
the angular difference $\psi =
\theta_1(\phi_2)-\theta_2(\phi_2)$ is expected to be large, and in this case,
we can approximate the
functions 
by the forms defined in (\ref{canonical2}).
When   $d_2/d_1$ is not close to  one (i.e. $d_2/d_1> 1.4$ 
according to experiments)
strong segregation effects act in the system.
In the case of thin flows treated in \cite{makse2}
(i.e. for  $R(x,t) \stackrel{<}{\sim}  3 * d_2$) 
there is no percolation 
in the rolling phase and 
all grains (large and small) interact with the surface, but 
the large grains are not being captured because of the large difference in
size. Thus the small grains are the only ones to be effectively interacting
with the sandpile surface resulting in a strong segregation effect.
If the flux of grains is large enough (i.e., for $R(x,t) \stackrel{>}{\sim} 
3 * d_2$, the
case treated here),
segregation  occurs in the rolling phase due to the percolation effect
so that the large grains
migrate to the top of the rolling phase and the small grains
migrate to the bottom of the rolling phase
\cite{savage1,drahun,makse3,cms}.
   Then small grains 
are the only ones interacting with the sandpile surface and again strong
segregation
is expected. This is confirmed by experiments done for $d_2/d_1> 1.4$
showing sharp segregation patterns when 
the large grains are smoother than the small grains 
$\delta\equiv \theta_{22}-\theta_{11} < 0$, and stratification 
when $\delta > 0$.

When the size ratio is  close to one ($d_2/d_1 < 1.4$),
the angle $\psi$ is expected to
be small and the collision 
functions  $a_\alpha$ and  $b_\alpha$
can be linearized around the respective angles of
repose $\theta_\alpha(\phi_2)$, since the region of interest 
is localized to a small region around the
angles of repose of the species. 
In this case segregation effects are expected to be less
important in comparison with the above case.
Indeed,  according to the experiments \cite{yan}
mixtures of grains differing in size and shape with
size ratio close to one 
 ($d_2/d_1 < 1.4$) give rise only to segregation and the concentration
show slow decay, i.e. the segregation pattern is not sharp as in the
above case.
The case of weak segregation 
is treated theoretically in \cite{bmdg}, where it is shown that 
the concentration profiles show slow (algebraic or power-law type) 
decay as a function
of the position in the cell.

\section{Profiles in the steady-state}
\label{ssregime}

We now calculate the steady-state solution of the equations of motion
for the two-species sandpile including the effect of percolation.

We consider the geometry of a silo
of lateral size $L$. We
assume that the difference 
$\psi = \theta_1(\phi_2) -
\theta_2(\phi_2) $
 is independent of the concentration $\phi_2$, then
$\psi = 
\theta_{11} - \theta_{21} = \theta_{12} - \theta_{22}$.
We set $v_1=v_2\equiv v$, and $\gamma_{11}=\gamma_{22}\equiv \gamma$.
We seek a solution  where the profiles of the sandpile
and of the
rolling grains do not change in time. Since stratification is an
oscillatory solution, stratification cannot be
observed for the steady-state solution. We set

\begin{mathletters}
\label{hr}
\begin{equation}
\frac{\partial h}{\partial t} = v ~ \frac{R^0}{L},
\label{h-steady}
\end{equation}
and 

\begin{equation}
\frac{\partial R_\alpha(x)}{\partial t}
= 0, 
\label{r-steady}
\end{equation}
\end{mathletters}
with the following boundary conditions:
\begin{eqnarray}
R_\alpha (x=0) = R_\alpha^0, & 
~~~~~~~~~~~~~~~~R_\alpha (x=L) =0,
\end{eqnarray}
with $R^0\equiv R_1^0+R_2^0$.

The profile of the total rolling species 
$R(x) = R_1(x) + R_2(x)$ is
\begin{equation}
R(x)=  \frac{ R^0}{L} ( L - x).
\label{total}
\end{equation}
which is obtained from (\ref{bdg-r}), (\ref{bdg-h}), 
and (\ref{hr}).
The linear profile for the total number of rolling species was also found 
in the case of one type of species in  \cite{pgg}. In fact this result 
is independent
of the type of interaction term used, and it is a consequence of the 
conservation of grains.

The equations for the rolling species are obtained from
(\ref{bdg-r}), (\ref{bdg-phi}), and
(\ref{hr}) as a function of the concentrations $\phi_\alpha$
\begin{equation}
\label{eq2}
{\displaystyle 
v \frac{\partial R_\alpha(x)}{\partial x} =
 -(v R^0/L) ~ \phi_\alpha(x),
}
\end{equation}

Since the collisions functions are defined according to the value
of the angle in comparison to the generalized angles of repose, 
we divide the calculations in two regions: Region A, where
  $\theta_2(\phi_2)<\theta<\theta_1(\phi_2)$, and  Region B, where 
  $\theta<\theta_2(\phi_2)<\theta_1(\phi_2)$.

{\bf Region A}. If $\theta_2<\theta<\theta_1$,
we obtain  
\begin{eqnarray}
\phi_1(x)=1,& ~~~~~~~~ \phi_2(x)=0.
\label{phi-up}
\end{eqnarray}
then we obtain the profile of the rolling species 
using (\ref{eq2}) and (\ref{phi-up})
\begin{equation}
\label{r1}
R_1(x) = R^0_1 -\frac{R^0}{L} x, \>\>\>\> 
R_2(x) = R^0_2.
\end{equation}
The profile of the sandpile is obtained from (\ref{bdg-h}),
(\ref{gamma}),
 and (\ref{h-steady})
\begin{equation}
\label{theta1}
\theta(x)-\theta_{11} 
=\frac{ -v/\gamma}{L R_1^0 /R^0 - x  }.
\end{equation}
This solution is valid when
$\theta(x)>\theta_1(\phi_2=1)=\theta_{21}$, 
or for  $x<x_m$, where $x_m$ is 
\begin{equation}
x_m = \frac{R_1^0}{R^0} L - \frac{v}{\gamma \psi}.
\end{equation}

We notice that in this region the solution is the same as the one found for
the case of thin flows without percolation ($\lambda = 0$)
in \cite{makse4}. This is because, even though the large grains 
may interact with the pile surface when percolation effects are absent, the 
small grains are captured before the large grains since the surface at the top
of the pile
is made of small grains, so that the large grains roll down very 
easily and they are not trapped (in other words, the angle of repose of
the pure small grains $\theta_{11}$ is larger than the angle of repose of 
the large grains rolling on a surface of small grains $\theta_{21}$).
The profile of the small rolling grains decays linearly with the distance as 
in the case of the total rolling species Eq. (\ref{total}), since essentially
the small grains are the only ones effectively interacting with the surface,
so that it is a single species problem.
Next, we show  that the percolation effect shows up in the region at the 
bottom of the pile causing corrections to the exponential 
profiles of the concentrations found in \cite{makse4}.

{\bf Region B}. 
If $\theta<\theta_2<\theta_1$, from (\ref{bdg-h}), (\ref{gamma}), 
and (\ref{h-steady}) 
we obtain

\begin{equation}
-\frac{v R^0}{\gamma L} = [\theta(x) - \theta_{1}(\phi_2)] R_1(x) + 
 [\theta(x) -
\theta_{2}(\phi_2)] R_2(x) e^{-\lambda R_1/R},
\end{equation}
and therefore
\begin{equation}
\label{t2}
\theta(x) - \theta_2(\phi_2) = \frac{-v R^0 /(\gamma L) + \psi
R_1(x)}{R^{*}(x)},
\end{equation}
where $R^{*}(x)=R_1(x) + R_2(x) e^{-\lambda R_1/R}$.
Using
  (\ref{t2})  we obtain the concentration as a function of
the rolling species

\begin{equation}
\label{c3}
{\displaystyle
\phi_1(x) = \frac{R_1(x)}{R(x)} \left ( 
1 + \frac{\gamma \psi L} {v R^0} R_2(x) e^{-\lambda R_1/R} \right ),
}.
\end{equation}
We obtain the equations for the rolling species
using Eqs. (\ref{eq2}) and  (\ref{c3}),

\begin{equation}
\label{q1}
{\displaystyle 
\frac{\partial R_1(x)}{\partial x} = -\frac{R_1(x)}{R^*(x)} \left(
\frac{ R^0}{L} + \frac{R_2(x)e^{-\lambda R_1/R}}{r} \right ),} 
\end{equation}
where  $r \equiv v/(\gamma \psi)$.
We set
\begin{equation}
u(x)=\frac{R_1(x)}{R^*(x)},
\end{equation}
and we obtain from (\ref{q1})
\begin{equation}
u'(x) = - u  \left(
-\frac{\partial R}{\partial x }\frac{1}{R^*(x)} + 
\frac{(1-u)e^{-\lambda u}}{r} \right ) - 
\frac{\partial R^*(x)}{\partial x} \frac{1}{R^*(x)}.
\end{equation}
This equation cannot be solved in closed form, but to a good approximation we
assume that $\partial R^*(x)/\partial x \approx \partial R(x)/\partial x$ (valid for small $\lambda$), 
and the equation is 
\begin{equation}
u'(x) \approx - u  
(1-u)e^{-\lambda u} /r,
\end{equation}
which has the solution
\begin{equation}
\frac{R_1(x)}{R^*(x)} \approx \frac{1}{1+C e^{-(x-x_m)/r}}
\end{equation}
where
 $C$ is an integration constant obtained
by considering the continuity at $x=x_m$. 
When there is no percolation effect $(\lambda=0)$
we recuperate the 
result found in \cite{makse4}, 
i.e. $R_1(x)/R(x) \sim \exp[-(x-x_m)/r]$ when $r\ll L$. 
Using Eq. (\ref{c3}) the concentration profiles can be obtained 
as a function of position.

\begin{figure}
\centerline{
\vbox{ \hbox{\epsfxsize=8.cm
\epsfbox{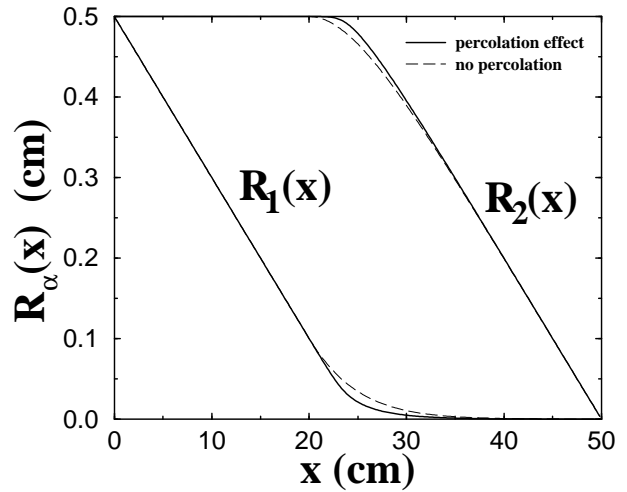}}
}}
\vspace{1cm}
\narrowtext
\caption{Results of the profile of rolling species when the 
percolation effect is present. Notice that small species (type 1) are
the only one interacting with the pile up to the center of the pile.
Then the large grains start to be captured to, and the number of
large rolling 
grains $R_2(x,t)$ decreases as well. For comparison we also plot in dashed line
the solution
corresponding to the case of thin flow where the percolation effect is not
acting. We see that the segregation in the case of percolation is stronger 
than the segregation in the case of no percolation effect.}
\label{results}
\end{figure}

To compare this solution with the case of no percolation effect, we obtain
the exact numerical solution of Eq. (\ref{q1}). 
Figure \ref{results} shows the solution we find for the rolling
species with percolation in the rolling phase in comparison with 
the solution corresponding to no percolation effect $\lambda = 0$.
We see that the percolation effect tends to increase the degree of
segregation of the mixture in comparison with the results found in
\cite{makse2} without percolation 
as expected;
the decay of the small rolling species when there
is segregation in the rolling phase is faster 
than the exponential decay found when there is no segregation in the rolling 
phase.

The concentration of small grains decays very fast near  the center of
the pile as observed in experiments \cite{makse3,yan}, with  a region of mixing with a characteristic size of  $r=v/\gamma
\psi$. 
For typical experimental values corresponding to a system of
$L=30$ cm, $d_1=0.27$ mm, $d_2=0.8$ mm we have \cite{makse3}
$v\simeq 10$ cm/sec, 
$\gamma \simeq 20$/sec, $R^0 \simeq 0.5$ cm, 
$\tan\psi=\tan \theta_{11} - \tan \theta_{21} \simeq 
0.2$ \cite{difference}.
Therefore, the region of mixing in the center of the pile
is of the order of $v/(\gamma \tan\psi) \simeq 2.5$ cm.

So far we have calculated the steady state solution of the problem. Numerical 
simulations \cite{makse2,cms,makse4,bmdg} shows that this type of solution
is only stable when the small grains are rougher than the large grains. 
When the large grains are the rougher, the solution turns unstable 
and the system 
displays the stratification pattern found in \cite{makse1}.

\section{Discussion}
\label{summary}

In summary, we 
study analytically  segregation  in granular
mixtures including the effect of segregation in the rolling phase, by
introducing a suitable modification of 
the interaction term in the BdG equations.
We find the corrections to the profiles of the steady-state of the
filling of a silo. The profiles show strong segregation effects in
agreement with experimental findings. The small corrections found here
for the case of thick flows of grains with percolation in the rolling phase
compared to the case
of thin flows but with a large difference in size ratio 
treated in \cite{makse4}
suggests the applicability of the 
results found in \cite{makse4} also 
to the case of thick flows.

ACKNOWLEDGMENTS.
I wish to thank T. Boutreux
for many stimulating discussions,  and BP and NSF for
financial support.

\end{multicols} 

\begin{references}
\bibitem{difference}
For comparison with experimental results,
all the values of the angles appearing in the theory
should be replaced by the 
tangent of the angle.
\end{references}
\end{document}